\documentclass[runningheads]{llncs}

\usepackage{graphicx}
\graphicspath{{./figures/}}
\usepackage{xcolor}
\usepackage{comment}
\usepackage{hyperref}

\usepackage[backgroundcolor=blue, textcolor=white]{todonotes}
\usepackage{listings}
\usepackage{color}
\usepackage{balance}
\definecolor{lightgray}{rgb}{.9,.9,.9}
\definecolor{darkgray}{rgb}{.4,.4,.4}
\definecolor{purple}{rgb}{0.65, 0.12, 0.82}

\lstdefinelanguage{JavaScript}{
  keywords={typeof, new, true, false, catch, function, return, null, catch, switch, var, if, in, while, do, else, case, break},
  keywordstyle=\color{blue}\bfseries,
  ndkeywords={class, export, boolean, throw, implements, import, this},
  ndkeywordstyle=\color{darkgray}\bfseries,
  identifierstyle=\color{black},
  sensitive=false,
  comment=[l]{//},
  morecomment=[s]{/*}{*/},
  commentstyle=\color{purple}\ttfamily,
  stringstyle=\color{red}\ttfamily,
  morestring=[b]',
  morestring=[b]"
}

\begin{document}
\title{The Rise of Disappearing Frameworks in Web Development}
%
%
\author{Juho Vepsäläinen\orcidID{0000-0003-0025-5540} \and Arto Hellas \and
Petri Vuorimaa}
\authorrunning{J. Vepsäläinen et al.}
%
\institute{Aalto University\\
\url{https://www.aalto.fi/en/department-of-computer-science}\\
\email{juho.vepsalainen@aalto.fi}}
\maketitle              
\begin{abstract}
The evolution of the web can be characterized as an emergence of frameworks paving the way from static websites to dynamic web applications. As the scope of web applications has grown, new technical challenges have emerged, leading to the need for new solutions. The latest of these developments is the rise of so-called disappearing web frameworks that question the axioms of earlier generations of web frameworks, providing benefits of the early web and simple static sites.


\keywords{web \and web development \and multi-page applications \and JavaScript \and front-end frameworks \and single-page applications \and islands architecture \and disappearing frameworks \and Astro \and Marko.js \and Qwik}
\end{abstract}
%
\subsubsection*{Acknowledgements}

This version of the contribution has been accepted for publication, after peer review (when applicable) at ICWE 2023 but is not the Version of Record and does not reflect post-acceptance improvements, or any corrections. The Version of Record is available online at: \url{https://doi.org/10.1007/978-3-031-34444-2_23}. Use of this Accepted Version is subject to the publisher’s Accepted Manuscript terms of use \url{https://www.springernature.com/gp/open-research/policies/accepted-manuscript-terms}

\section{Introduction} \label{sec:introduction}

The World Wide Web (WWW) was designed to bring a global information universe into existence using the prevailing technology~\cite{bernerslee1992}. Over time, it grew into a whole application platform. The evolution into a platform could be characterized through the following phases: (1) the birth of the web in the early 1990s~\cite{bernerslee1992}, (2) the introduction of AJAX in 1999~\cite{flanagan1998}, and (3) rise of the Single-page applications (SPAs) starting from the early 2010s. Each of the phases was a response to the rising need for interactivity and improvements made to the browsers. For example, the introduction of AJAX let developers manipulate server data without having to refresh the whole page shown to the client. This led to more interactive experiences and, eventually, SPAs, which dominate today's dynamic web applications. As a response to SPAs, a new breed of frameworks, so-called \emph{disappearing frameworks}, have emerged.

The purpose of this article is to examine (RQ1) why disappearing frameworks are needed, (RQ2) what disappearing frameworks are, and (RQ3) why disappearing frameworks might be the next major phase in web development. To address the first research question RQ1, we'll discuss the background leading to the current situation in Section~\ref{sec:background}. To understand the concept of disappearing frameworks -- RQ2 -- in detail, we'll delve into the concept and a couple of examples in Section~\ref{sec:disappearing-frameworks}. In Section~\ref{sec:discussion} we argue why disappearing frameworks might become the next major phase in web development, addressing RQ3.

\section{Background} \label{sec:background}

Originally the web was designed as a content platform. Over time, as its usage grew, so did the user requirements for interactivity. As a result, it has morphed into an application platform spanning from desktop to mobile devices and beyond. Web technologies have become a universal way to deliver software to countless users. In this section, we'll have a brief look at the major movements and their contribution as this will help us to understand the motivation behind the development of disappearing frameworks.



\subsection{First decades of the web}

Early websites of the 1990s were authored directly in key web technologies: HTML, CSS, and JavaScript. These technologies are still in use and form the building blocks of the web as we know it. Authoring websites has however changed as contents may be generated either via code or via a user interface, depending on the system. The idea of generating code is not novel, as editors such as Frontpage and Dreamweaver offered the opportunity to create websites in a graphical manner already in the 1990s.

To manage a shared state, server-side functionality implemented using technologies such as PHP~\cite{lerdorf2002} and cgi-bin~\cite{cgibin} was used. These allowed the implementation of complex logic on the server side, which could be customized to fit the need of individual organizations. As the needs of organizations were often related to editing, storing, sharing, and displaying content, Content Management Systems (CMSs) such as WordPress emerged. Slowly, CMSs came to dominate the content market~\cite{w3techs}, providing benefits for both developers and users~\cite{benevolo2007,boiko2005,yermolenko2021}.

The early web applications were most often Multi-page applications (MPAs), where the application state lives on the server, and each request from the client to the server loads a new page \cite{kaluvza2018comparison}. The introduction of JavaScript in 1995~\cite{flanagan1998} and in particular Asynchronous JavaScript (AJAX) in 1999 allowed developers to move logic to the client over time, which led to the birth of JavaScript-heavy programming models such as Single Page Applications (SPAs).


\subsection{The rise of single-page applications}

SPAs were developed to address the constraints of MPA development, mainly the need to refresh the page when data changes \cite{kaluvza2018comparison}. AJAX was the key enabler as a technology as it allowed developers to retrieve data from the server. With the introduction of the History API in HTML5 specification in 2008, it became possible to control routing as well. To understand the differences between MPAs and SPAs, we've listed their relative merits in Table~\ref{table:mpavsspa}, summarizing~\cite{iskandar2020comparison,kaluvza2018comparison,solovei2018difference}.

\begin{table}[ht]
 \caption{MPA vs. SPA}
\begin{tabular}{ |p{3.5cm}|p{2.5cm}|p{6cm}| }
 \hline
 Dimension & MPA & SPA\\
 \hline
 Relies on JavaScript & No & Yes \cite{kaluvza2018comparison,solovei2018difference}\\
 Initial cost of loading & Potentially low & High due to dependency on JavaScript \cite{solovei2018difference} \\
 Overall response time & Slower & Faster due to partial updates \cite{kaluvza2018comparison}\\
 Business logic & Coupled & Decoupled \cite{kaluvza2018comparison,solovei2018difference}\\
 Refresh on navigation & Yes & No \cite{solovei2018difference}\\
 Bandwidth usage & Higher & Lower due to only transaction-related data moving between the parties \cite{kaluvza2018comparison} \\
 Offline support & Not possible & Possible \cite{kaluvza2018comparison,solovei2018difference}\\
 Search Engine Optimization (SEO) & Excellent & Possible but difficult \cite{iskandar2020comparison,kaluvza2018comparison,solovei2018difference}\\
 Security & Understood & Practices still being established \cite{kaluvza2018comparison} \\
 Routing & At server & Duplicated in server and client \cite{solovei2018difference}, but modern frameworks, such as Next.js, mitigate the problem \\
 \hline
\end{tabular}
 \label{table:mpavsspa}
\end{table}

SPAs set a new standard for Developer eXperience (DX) and what's expected from developer tooling. The developer-side improvements come with a cost as SPAs direct developers to rely on JavaScript and ignore native browser APIs, which have improved over time~\cite{ryan2021}. Techniques, such as progressive enhancement~\cite{alistapart2008} can help, but a more significant change is required on the tooling level to address the user needs~\cite{ryan2021}.

\subsection{The current front-end development landscape}

The current front-end development landscape is dominated by React, Vue, and Angular\footnote{Based on Stack Overflow Survey of 2022 \cite{soSurvey2022}, 42.6\% of developers use React. For Angular, the proportion was 20.4\%, and for Vue 18.8\%. W3Techs \cite{w3techsReactVuejs2022} puts these values into perspective as there the global market share of React is around 3.7\% for all websites.}. Out of these technologies, React and Vue can be considered libraries in the sense that they have a strict focus and they omit opinions such as how routing should be implemented. Angular can be considered a framework as it comes with everything one would need to construct an application. Due to this, frameworks such as Next.js or Nuxt.js\footnote{Based on \cite{soSurvey2022}, 13.5\% of developers use Next.js so that means roughly one out of four React developers use Next.js. For Nuxt.js, the figure is 3.8\% which means roughly one out of six.} have emerged as they provide structure and opinionated development practices on top of existing libraries. Regardless of the library or framework, the contemporary solutions rely on the three following  key principles:

\begin{enumerate}
    \item Component orientation -- They come with component-oriented abstractions\footnote{By component-oriented abstractions, we mean a way to encapsulate markup and potentially local state so that it can be reused.}
    \item Templating -- A solution, such as JSX, is used for modeling markup
    \item Hydration -- To make the components come alive, the solutions run the code at the client side. By this, we mean attaching event handlers and running component logic \cite{huotala2021benefits}.
\end{enumerate}

Each library and framework relies on some amount of client-side JavaScript to load a web page -- that is, when a website is loaded, client-side JavaScript is executed to display content. \href{https://preactjs.com/}{Preact} and \href{https://github.com/vuejs/petite-vue}{petite-vue} are notable examples of implementations that try to minimize the size of the required JavaScript while keeping API compatibility with their bigger brothers.


\subsection{A need to reconsider technologies}

Given the current solutions allow developers to build complex and dynamic web applications, what is there left to do? As pointed out in~\cite{jason2019}, analyzing the characteristics of real-world applications is difficult. It is difficult to generalize and extract best practices as they are context-specific. In a case where an MPA might be a good solution, a SPA could fare poorly. To capture and characterize specific use cases, ~\cite{jason2019} proposes a set of holotypes that can be used to make informed decisions on deciding what sorts of technical approaches one could or should make.


The key differences between the holotypes proposed in~\cite{jason2019} relate to the following dimensions: (1) interactivity, (2) search engine optimization (SEO), (3) media type (rich or not), (4) client size (thick or not), (5) processing (server or client), (6) latency, (7) offline-capability. This list summarizes the complexity related to modern web development and associated technical choices.

The list also highlights a need to reconsider technologies used for web development. Rather than trying to fit solutions such as SPAs to various use cases, \emph{disappearing frameworks give the option to reframe the problems and to potentially address them from an angle that could be considered a paradigm-level shift} (RQ1). This is the foundation of why we consider them as the next potential phase in mainstream web development.

\section{Disappearing frameworks} \label{sec:disappearing-frameworks}

Disappearing frameworks start from close to zero cost in terms of JavaScript loaded by the client and remove it from the application as much as possible~\cite{ryan2021}\footnote{The idea of disappearing frameworks is also in line with Transitional Web Applications (TWAs), where the key idea is to draw from both the traditional web and SPAs~\cite{ryan2021,rich2021} to build applications with the following characteristics: (1) static site rendering is utilized for fast initial loading times (2), resiliency is achieved by allowing applications to work without JavaScript by default, and (3) consistent experience and accessibility are built-in by definition~\cite{ryan2021}}. The shift in focus provides a strong contrast to the current mainstream JavaScript frameworks, which come with a heavy upfront cost that increases as components are added to the system~\cite{solovei2018difference}. Due to this shift in focus, applications are distributed into smaller sections instead of treating the whole as a single top-down system. The approach works well with MPAs as it allows developers to pull from experiences gained through building traditional systems while gaining modern DX as experienced with SPAs. As a side effect, the ideas of SPAs and MPAs may eventually merge and become a framework-level concern allowing the developers to use what is best for the given situation. Here, we discuss the key features of disappearing frameworks: (1) islands architecture and (2) automatic islands, outlining an answer to (RQ2).

\subsection{Islands architecture}

Islands architecture is the first stepping stone toward disappearing frameworks. It is an approach that can be considered further evolution of progressive enhancement. In the islands architecture, web page sections are treated as separate islands, some of which might be completely static while some might have some interactivity in them \cite{jason2020}. A good implementation of islands architecture should allow the developer to decide when an interactive island is loaded \cite{patterns2022}. Controlling loading is important as it lets the developers defer secondary work or even avoid it. Deferring work is the key to improving performance and decreasing bandwidth usage, which is important e.g., in mobile contexts and search engine optimization. 

\begin{figure}
    \centering
    \includegraphics[scale=0.23]{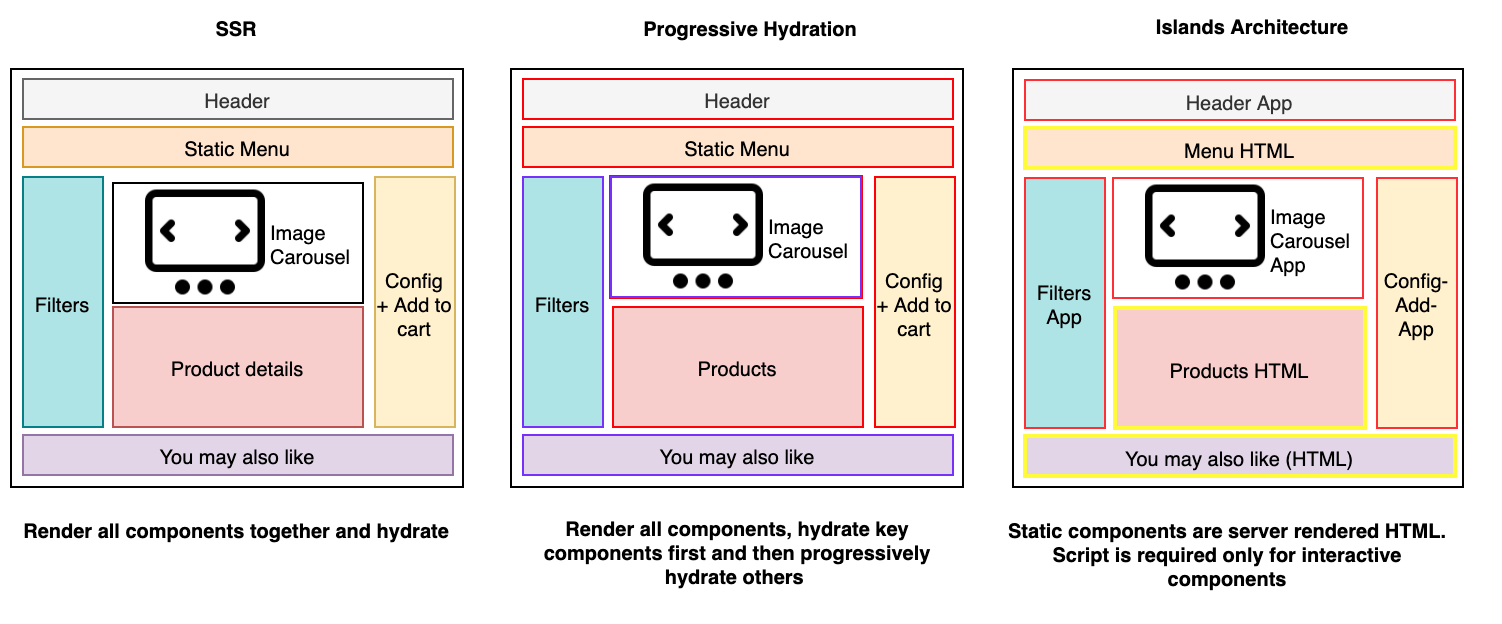}
    \caption{Islands architecture compared \cite{patterns2022}}
    \label{fig:islands-compared}
\end{figure}

Fig. \ref{fig:islands-compared} compares the islands architecture to other contemporary architectures, namely SSR and progressive hydration. In progressive hydration, the application framework controls which portions of a page to hydrate~\cite{patterns2022}. In islands architecture, the concern is distributed so that each island can load independently from the others in an asynchronous fashion~\cite{patterns2022}. Therefore, the difference between the approaches has to do with control.

The benefits of islands architecture include (1) avoidance of top-down rendering \cite{jason2020}, (2) improved performance as there is less JavaScript to ship \cite{jason2020,patterns2022}, (3) less problematic SEO story due to an increased amount of static content \cite{jason2020}, (4) better accessibility and discoverability by default~\cite{patterns2022}, and (5) component-orientation to encourage reuse~\cite{patterns2022}. The last point is not without problems as usage of islands can lead to composition issues since it is not trivial to compose functionality across an island boundary. Also, the SEO point comes with caveats related to SPAs and is not without complexity.

The islands architecture has been criticized since as a new approach, official support is limited~\cite{patterns2022}, there's limited discussion around the idea (besides ~\cite{jason2020,patterns2022,etsy2021}), and frameworks might claim to implement the idea while implementing it wrong~\cite{patterns2022}. In addition, the architecture might not fit a highly interactive use case such as a social media application with a high amount of interactivity~\cite{patterns2022}.

Given the islands architecture has been formalized only in 2019~\cite{jason2020}, there is no widespread usage yet. According to Hallie and Osmani~\cite{patterns2022}, at least the following frameworks support it: \href{https://astro.build/}{Astro}, \href{https://www.11ty.dev/}{Eleventy} when combined with \href{https://preactjs.com/}{Preact}. Out of these, only Astro was developed with the architecture in mind, whereas the other options contain the features needed for a developer to implement it. Further, Astro is not coupled to a specific library, such as React, but instead lets the developer choose and even mix solutions based on the need at hand.

\subsection{Compiler-based approaches – Marko.js and Qwik}

Both \href{https://markojs.com/}{Marko.js} and \href{https://qwik.builder.io/}{Qwik} implement a compiler-based approach that is able to write split points for code automatically. In Qwik's case, the compiler is able to split code at the view, state, and handler level~\cite{devYourBundler}. Marko.js generates split points at the sub-component level as well~\cite{fluurt2020}. It is these split points that represent code that can be loaded later on demand. In earlier approaches, such as React, the developer would have to insert hints\footnote{The latest ECMAScript standard supports this in the form of \textit{import()} statement.} for the bundler in the code to tell which part of code can be loaded later.


Marko.js is a project originating from eBay, and based on its revision history, its development has begun as early as 2014. Qwik is a younger project whose earliest commit can be traced back to early 2021. Whereas Marko.js is characterized as a language in its communications, Qwik is web framework that addresses the conflicting requirements of interactivity and page speed \cite{devmQwiksMagic}. The conflict means you have to compromise in either interactivity or speed due to the hydration cost of the current frameworks \cite{devmQwiksMagic}.

At the end of 2020, Marko.js changed its direction as the project pivoted to rework its engine to (1) reduce shipped JavaScript size, (2), improve client-side performance, and (3) improve development experience~\cite{fluurt2020}. The ideas are consistent with the ideas behind disappearing frameworks and show that client-side experiences have to be increasingly interactive yet light.


Qwik solves compromising interactivity and speed by choosing resumability over hydration, automatic code splitting, tiny runtime, and compilation over runtime cost~\cite{qwikdocs}. Presently, these factors make Qwik unique when compared to options. Qwik is consistent with the idea of disappearing frameworks as it tries to make itself vanish from the application. Because of its approach, Qwik represents a paradigm-level shift in how to develop web applications. One of the ways Qwik achieves its targets is by focusing on different metrics than its predecessors, namely Time to Interactive (TTI) over Time to Load (TTL).










\section{Discussion and Conclusion} \label{sec:discussion}

To summarize, the growth of the web has been fueled by increasing requirements from both developers and users. Early web practices, including MPAs, were later complemented by developer-friendly approaches, such as SPAs. During the process, fundamental ideas of the web were put aside as applications required JavaScript to run and came with accessibility and SEO challenges. We discussed disappearing frameworks and argue that they represent the next major stage in web development as they take the best parts of SPAs and combine them with the good development practices of the early web, improving common challenges of SPAs such as search engine optimization. The existing early implementations, such as Marko.js and Qwik, show promise and the increasing popularity of Astro highlights the demand for lighter ways to develop for the web; it also demonstrates that there is room for more than one option in the ecosystem, as it provides a framework/library agnostic approach to web development. These possibilities, coupled with developer experience, lead to us believe that they might be the next major phase in web development (RQ3).

By questioning existing axioms, the emergence of disappearing frameworks leads to a group of new research questions. Given the frameworks postpone loading code, what is the optimal strategy to do it? What are the pros/cons of the solutions from a developer and a user perspective relative to the incumbent approaches? What is the cost of using the island architecture? What is the cost of using hydration based technologies in islands? Is the idea of disappearing frameworks compatible with other rising approaches, such as micro-frontends? How does the islands architecture and disappearing frameworks scale as application size grows?

Disappearing frameworks provide a refreshing take on web development as they promise benefits for both developers and users. At the same time, there are many questions related to the feasibility of the approach as no case studies exist.

\bibliographystyle{splncs04}
\bibliography{references}

\begin{thebibliography}{10}
\providecommand{\url}[1]{\texttt{#1}}
\providecommand{\urlprefix}{URL }
\providecommand{\doi}[1]{https://doi.org/#1}

\bibitem{benevolo2007}
Benevolo, C., Negri, S.: Evaluation of content management systems (cms): a
  supply analysis. Electronic Journal of Information Systems Evaluation
  \textbf{10}(1),  pp9--22 (2007)

\bibitem{bernerslee1992}
Berners-Lee, T., Cailliau, R., Groff, J.F., Pollermann, B.: World-wide web: the
  information universe. Internet Research  (1992)

\bibitem{boiko2005}
Boiko, B.: Content management bible. John Wiley \& Sons (2005)

\bibitem{fluurt2020}
Carniato, R.: {F}{L}{U}{U}{R}{T}: {R}e-inventing {M}arko --- dev.to.
  \url{https://dev.to/ryansolid/fluurt-re-inventing-marko-3o1o} (2020),
  [Accessed 11-Jan-2023]

\bibitem{ryan2021}
Carniato, R.: Understanding transitional javascript apps (Nov 2021),
  \url{https://dev.to/this-is-learning/understanding-transitional-javascript-apps-27i2},
  [Accessed 29-Sep-2022]

\bibitem{cgibin}
Common {Gateway} {Interface} (Aug 2022),
  \url{https://en.wikipedia.org/w/index.php?title=Common_Gateway_Interface&oldid=1102228140},
  page Version ID: 1102228140

\bibitem{flanagan1998}
Flanagan, D., Novak, G.M.: Javascript: The definitive guide (1998)

\bibitem{alistapart2008}
Gustafson, A., Overkamp, L., Brosset, P., Prater, S.V., Wills, M., PenzeyMoog,
  E.: Understanding progressive enhancement (Oct 2008),
  \url{https://alistapart.com/article/understandingprogressiveenhancement/},
  [Accessed 29-Sep-2022]

\bibitem{patterns2022}
Hallie, L., Osmani, A.: {I}slands {A}rchitecture --- patterns.dev.
  \url{https://www.patterns.dev/posts/islands-architecture/} (2022), [Accessed
  29-Sep-2022]

\bibitem{rich2021}
Harris, R.: Have single-page apps ruined the web? | transitional apps with rich
  harris, nytimes (Oct 2021),
  \url{https://www.youtube.com/watch?v=860d8usGC0o}, [Accessed 29-Sep-2022]

\bibitem{devYourBundler}
Hevery, M.: {Y}our bundler is doing it wrong --- dev.to.
  \url{https://dev.to/builderio/your-bundler-is-doing-it-wrong-ic0} (2021),
  [Accessed 14-Nov-2022]

\bibitem{huotala2021benefits}
Huotala, A.: Benefits and Challenges of Isomorphism in Single-Page
  Applications: A Case Study and Review of Gray Literature. Master's thesis,
  University of Helsinki (2021)

\bibitem{iskandar2020comparison}
Iskandar, T.F., Lubis, M., Kusumasari, T.F., Lubis, A.R.: Comparison between
  client-side and server-side rendering in the web development. In: IOP
  Conference Series: Materials Science and Engineering. vol.~801, p. 012136.
  IOP Publishing (2020)

\bibitem{etsy2021}
Jones, A.: Etsy engineering: Mobius: Adopting jsx while prioritizing user
  experience.
  \url{https://www.etsy.com/codeascraft/mobius-adopting-jsx-while-prioritizing-user-experience/}
  (2021), [Accessed 29-Sep-2022]

\bibitem{kaluvza2018comparison}
Kalu{\v{z}}a, M., Troskot, K., Vukeli{\'c}, B.: Comparison of front-end
  frameworks for web applications development. Zbornik Veleu{\v{c}}ili{\v{s}}ta
  u Rijeci  \textbf{6}(1),  261--282 (2018)

\bibitem{lerdorf2002}
Lerdorf, R., Tatroe, K., Kaehms, B., McGredy, R.: Programming Php. O'Reilly
  Media, Inc. (2002)

\bibitem{jason2019}
Miller, J.: {A}pplication {H}olotypes: {A} {G}uide to {A}rchitecture
  {D}ecisions - {J}{A}{S}{O}{N} {F}ormat --- jasonformat.com.
  \url{https://jasonformat.com/application-holotypes/} (2019), [Accessed
  10-Jan-2023]

\bibitem{jason2020}
Miller, J.: Islands architecture (2020),
  \url{https://jasonformat.com/islands-architecture/}, [Accessed 29-Sep-2022]

\bibitem{qwikdocs}
Qwik: {O}verview - {Q}wik --- qwik.builder.io.
  \url{https://qwik.builder.io/docs/overview/} (2022), [Accessed 14-Nov-2022]

\bibitem{devmQwiksMagic}
{Q}wik's magic is not in how fast it executes, but how good it is in avoiding
  doing any work --- devm.io.
  \url{https://devm.io/javascript/qwik-javascript-hevery} (2022), [Accessed
  15-Nov-2022]

\bibitem{w3techsReactVuejs2022}
{R}eact vs. {V}ue.js vs. {A}ngular usage statistics, {O}ctober 2022 ---
  w3techs.com.
  \url{https://w3techs.com/technologies/comparison/js-angularjs,js-react,js-vuejs}
  (2022), [Accessed 31-Oct-2022]

\bibitem{solovei2018difference}
Solovei, V., Olshevska, O., Bortsova, Y.: The difference between developing
  single page application and traditional web application based on mechatronics
  robot laboratory onaft application. Automation of technological and business
  processes  \textbf{10}(1) (2018)

\bibitem{soSurvey2022}
{S}tack {O}verflow {D}eveloper {S}urvey 2022 --- survey.stackoverflow.co.
  \url{https://survey.stackoverflow.co/2022/}, [Accessed 31-Oct-2022]

\bibitem{w3techs}
{W}3{T}echs - extensive and reliable web technology surveys.
  \url{https://w3techs.com/} (2022), [Accessed 03-Oct-2022]

\bibitem{yermolenko2021}
Yermolenko, A., Golchevskiy, Y.: Developing web content management
  systems--from the past to the future. In: SHS Web of Conferences. vol.~110.
  EDP Sciences (2021)

\end{thebibliography}
\end{document}